\documentclass[10pt,conference]{IEEEtran}

\usepackage{booktabs}
\usepackage{array}
\usepackage{cite}
\usepackage{color,soul}
\usepackage{amsmath,amssymb,amsfonts}
\usepackage{algorithm}
\usepackage[noend]{algpseudocode}
\usepackage{graphicx}
\usepackage{textcomp}
\usepackage{xcolor}
\usepackage{comment}
\newcommand{\link}[1]{#1}

\usepackage{listings}
\def\BibTeX{{\rm B\kern-.05em{\sc i\kern-.025em b}\kern-.08em
    T\kern-.1667em\lower.7ex\hbox{E}\kern-.125emX}}

\graphicspath{ {./figures/} }

\begin{document}
\title{Quartermaster: A Tool for Modeling and Simulating System Degradation}

\author{\IEEEauthorblockN{Matt Pope, Jonathan Sillito}
  \IEEEauthorblockA{
    Brigham Young University\\
    Provo, Utah\\
    mattpope@byu.edu, sillito@byu.edu}}

\maketitle

\begin{abstract}
  It is essential that software systems be tolerant to degradations in components they rely on. There are patterns and techniques which software engineers use to ensure their systems gracefully degrade. Despite these techniques being available in practice, tuning and configuration is hard to get right and it is expensive to explore possible changes to components and techniques in complex systems. To fill these gaps, we propose Quartermaster to model and simulate systems and fault-tolerant techniques. We anticipate that Quartermaster will be useful to further research on graceful degradation and help inform software engineers about techniques that are most appropriate for their use cases.
\end{abstract}

\section{Introduction}


Software systems are often comprised of multiple interdependent components. A failure or degradation in one component can lead to failures in dependent components and even whole system failure. To avoid such cascading failures, systems may be engineered to tolerate component failures~\cite{White_1996} or to gracefully degrade by reducing functionality or performance without failing completely~\cite{Shelton_2003}. To this end, engineers use a range of fault-tolerant \cite{White_1996} techniques~\cite{Nygard_2007, Livora_2017, Heorhiadi_2016} to implement graceful degradation such as circuit breakers, retry strategies~\cite{Chan_2007, Zheng_2010}, priority queues~\cite{Ellens_2012}, load shedding~\cite{Beyer_2016}, and caching. Many systems use some combination of techniques rather than relying on just a single technique. 


In our experience, 
there are several common problems that arise when trying to use these techniques. It is expensive, and sometimes impractical, to explore which techniques are right for production systems since it often requires setting up duplicate environments and replicating live traffic in the system. Tuning the selected techniques is often difficult to get right, requiring expertise and extensive testing to configure~\cite{Janardanan_2015}. 
Despite these challenges, exploration plays an important role in engineering a system capable of tolerating failures.

Quartermaster\footnote{https://github.com/BYU-SE/quartermaster} is a new modeling and simulation tool for engineers and researchers to explore (existing and novel) fault-tolerant techniques and their effects on a system's behavior under different scenarios. With Quartermaster, a user can create a model of a software system of interest, including the fault-tolerant techniques used in the system, by writing TypeScript code or using predefined components that Quartermaster provides. A user can then simulate the execution of that model under various scenarios (e.g., a surge in traffic) and review the output to understand how the system behaved. This inexpensive exploratory process can continue as the model and scenarios are modified and the simulation is run again, aiding users in improving a system's ability to tolerate failure.

The goal of this paper is to introduce Quartermaster and demonstrate its usefulness. To that end, we first describe how Quartermaster can be used, by working through an example (see Section~\ref{example_use_case}). We next present in more detail how to model a system and individual fault-tolerant techniques in Quartermaster (see Section~\ref{model}) and how to use Quartermaster to simulate executions of such models (see Section~\ref{simulating_with_quartermaster}). Finally, we report on a partial evaluation of Quartermaster (see Section~\ref{evaluation}).

\section{Example Use Case}
\label{example_use_case}

To demonstrate the use and usefulness of Quartermaster, we will discuss how it could be used by engineers in the context of a software engineering scenario. The scenario we have selected is based on an actual system failure from 2015, published in a publicly available incident report.\footnote{https://circleci.statuspage.io/incidents/hr0mm9xmm3x6}

The system that failed is called CircleCI and the parts of the system related to the failure include a queue that receives events from an external source, and a service that reads events from the queue and queries a database. The failure was precipitated by an unusually large number of incoming events to the queue, over a short period of time. This surge in events led to a large number of concurrent database queries causing the database to become overloaded. Event throughput dropped and the queue continued to grow in size for several hours until CircleCI engineers mitigated the incident.


After the failure was mitigated, the engineers explored various architectural and configuration changes, or other optimizations to make their system able to tolerate potential future event surges. Fundamentally, helping engineers with such explorations is the goal of Quartermaster; either based on an existing system and incident as we are illustrating here, or based on a planned system and anticipated events.

Making changes to a production system and performing appropriate testing (e.g., load testing) is expensive and time consuming, limiting the scope of possible explorations. In contrast, exploration in Quartermaster requires minimal code and time allowing for a more comprehensive exploration of the space of possibilities. That is not to say that Quartermaster eliminates the need to implement and test changes to the system, but it does allow engineers to more easily explore a larger set of options before deciding what to implement and test in the actual system.

To use Quartermaster for this scenario we first implemented a \emph{model} of the CircleCI system, or specifically the parts of their system that are relevant to the incident, as a series of three stages: an event processing stage (which includes the event queue), a stage modeling the retry strategy used by CircleCI, and a database stage. Implementing these stages required 76 lines of code. This code is available in a Github repository\footnote{https://github.com/BYU-SE/idb-quartermaster-models} and discussed further in Section~\ref{model}.


We next ran our simple CircleCI model, simulating a surge of events to the queue and confirmed that our model failed in the same way the actual system failed during the 2015 incident we discussed above. Simulating that event surge involved specifying a probability distribution for the arrival of events and a few other parameters, which together define the \emph{scenario} for a run of the simulator. For more details on running simulations in Quartermaster, see Section~\ref{simulating_with_quartermaster}.

With this model in place, we next began considering changes that might improve CircleCI's ability to deal with a surge of events. The process we followed, and the process we expect others to follow when using Quartermaster, begins with an idea (for something to change) or equivalently a question and then follows these steps: (1) modify the model of the system or its configuration details, (2) specify a scenario to be used by the framework in the simulation's execution, (3) execute the simulation and observe the resulting system behavior, and (4) repeat as necessary. 

As an illustration, we followed that iterative process to explore the effects of (A) bounding the size of the queue, (B) limiting the number of queue workers (i.e., the number of threads reading from the queue and calling the database), and (C) eliminating the retry strategy, in the context of an event surge. Each of these models are variations on the original model requiring simply changing one or two lines of code and each simulation takes just a few minutes to execute.

A summary of the results of our simulations are listed in Table~\ref{tbl:explorations}. The metrics we are reporting include the event rejection rate (percent of events that are rejected by the queue), availability (percent of events that succeed), maximum number of events in the queue over the course of the simulation, mean queue wait time for both successful and unsuccessful events, mean event throughput (i.e., mean event processing time), and finally the time it took the system to recover after the surge ended (how quickly it was able to process the surge in events completely). The way time is measured in Quartermaster will be described in Section~\ref{simulating_with_quartermaster}.

\begin{table}[t]
  \centering
  \caption{Results from simulating an event surge for the original model (at the time of the incident) and three variations.}
  \label{tbl:explorations}
  \begin{tabular}{lrrrr}
  \toprule
  
  {\bf Metric} &
  {\bf Original} &
  {\bf Model A} &
  {\bf Model B} &
  {\bf Model C} \\
  \midrule
  
  Event rejection rate          & 0\%       & 79\%      & 0\%        & 0\%      \\
  \addlinespace[.1em]
  Availability                & 91\%      & 19\%      & 100\%     & 70\% \\
  \addlinespace[.1em]
  
  Max queue Size          & 49,562    & 10,000    & 36,063    & 49,472 \\
  \addlinespace[.1em]

  Mean time in queue      & 3,206     & 132       & 15        & 2,459 \\
  \hspace{0.75em}Success events  & 3,201     & 633       & 17        & 2,466 \\
  \hspace{0.75em}Failed events     & 3,255     & 15        & -         & 2,440 \\
  \addlinespace[.1em]
  

  Throughput              & 0.00769   & 0.0373    & 1.2311    & 0.0100 \\
  \addlinespace[.1em]
  Recovery time           & 6,498     & 1,340     & 41        & 4,995 \\
  
  \bottomrule
  \end{tabular}
  \vspace{-0.197in}
\end{table}

As the particular values for each of the metrics listed in the table are sensitive to the precise configuration used for the execution, we expect the relative values and the overall behavior to be more informative than the absolute values. The results show that model A (which used a bounded queue) sheds load quickly and reduces total work being done by the system at the cost of a high rejection rate. For model B (which limited the number of queue workers), despite the large increase of traffic over a short period of time, there is no drop in throughput and it recovers quickly, though determining the ideal number of workers to maximize the system throughput without degrading requires tuning, best performed on the live system. Model C (which eliminated retries) shows that you can trade time in the queue for an increased chance of a failure. Understanding the desirability of the tradeoffs represented by these three models requires business understanding we do not have and is outside the scope of this short paper. 

These are simple explorations and are related to common fault-tolerant techniques, such as load leveling and load shedding, but they illustrate the use of Quartermaster in understanding how various fault-tolerant techniques affect the behavior of a system.

\section{Quartermaster Model}
\label{model}

In Quartermaster both system components and techniques used to mitigate or protect against failures in those components are modeled as \emph{stages}. A stage is a unit of event processing, or to put it another way, an object that can handle events. An event represents a generic unit of work to be done by the system and what that work is will depend on the domain of the system being modeled. For example, when modeling a web application, each event could represent an HTTP request.

During a simulation, event handling within a stage follows a simple flow and each stage defines event handler methods that allow a stage to customize how it handles events. This event handling can be understood as following two steps:

(1) \emph{Event admission} is controlled by a stage's $add()$ method. When an event is passed to the stage, the stage can accept or reject that event before processing the event.

(2) \emph{Event processing} is defined by a stage's $workOn()$ method. This work may be synchronous, such as a retry which waits for a failure before attempting again. It may be asynchronous, such as a cache that returns immediately and then queries a key refresh in a background thread. This concurrency model creates the flexibility to represent a variety of implementations for systems and fault-tolerant techniques.

When modeling an existing or planned system in Quartermaster, one system component or one fault-tolerant technique may be mapped to one stage or multiple stages. Alternatively, multiple components may be mapped to one stage. Quartermaster is unopinionated about this decomposition, and so the level of granularity needed will depend on the behavior the user wants to include in the model.

Instead of modeling an entire system, we expect that users of Quartermaster will model specific components or techniques of a system, and many of those will be greatly simplified. For example, when we modeled the CircleCI system, as described in Section~\ref{example_use_case}, we used a network of three stages representing two system components and a retry technique between these two components. We defined custom database and build service stages in addition to a retry stage that is one of a handful of predefined stages provided by Quartermaster. The database stage was represented simply as a latency distribution (which we sample from using the $exponential()$ function in the code below) and an error distribution, without any internal application logic. Lastly, we use a simple queue to limit the number of concurrent events being handled by the database. Here is a simplified version of the database stage. The complete code for this example stage is available in the repository for the Quartermaster tool. 

{\small
\begin{lstlisting}
class Database extends Stage {
  public availability:number = 0.9995;
  constructor() {
    super();
    this.inQueue = new FIFOQueue(1, 300);
  }
  async workOn(event: Event): Promise<void> {
    const latency = exponential(...);
    await metronome.wait(latency);

    if(Math.random() > this.availability) 
      throw "fail";
  }
}
\end{lstlisting}}


We connected those three stages using the following code, with the result that the retry stage wraps the database stage and the build service stage wraps the retry stage. 

{\small
\begin{lstlisting}
const db = new Database();
const retry = new Retry(db);
const service = new BuildService(retry);
\end{lstlisting}}

The Quartermaster model allows a user to characterize and compare existing techniques, such as retry strategies, caching, and timeouts, and we hope it will promote the discovery of novel techniques. We also expect the simplicity of modeling to appeal to engineers who are interested in exploring various models in various scenarios as a preliminary investigation before making and testing changes to a production system.

\section{Simulating with Quartermaster}
\label{simulating_with_quartermaster}

Once a model of a system has been created, Quartermaster can simulate the execution of that model under various scenarios. The output of the simulated execution allows a user to understand how the model behaved in the given scenario. In this section, we describe how to define a scenario, how to configure the output or reporting, and finally how to simulate an execution.

To understand these parts, it is important to understand that the discrete-event simulation coordinates the activity of all stages using an abstract notion of time referred to as a \emph{tick}. This helps decouple the simulation from hardware and software limitations and allows users to choose a comparable real-world unit. In this paper, a tick can be considered equivalent to a millisecond, and 1,000 ticks equivalent to a second.

A scenario in Quartermaster is the contextual information used to drive the simulation, and it can change over the course of the simulation. At a minimum, defining a scenario involves specifying: (1) the rate of event arrival, expressed as a number of events per 1,000 ticks, and (2) parameters for the keyspace, which defines the distribution of unique requests without having to describe the complexity of an actual request. For example, while simulating the CircleCI event surge, we created the scenario using the following code, which defined a keyspace using a discrete normal distribution with a mean of 1,000 and a standard deviation of 200 and provided an initial event rate and then a steady increase. 

{\small
\begin{lstlisting}
simulation.keyspaceMean = 1000;
simulation.keyspaceStd = 200;
simulation.eventsPer1000Ticks = 1500;
metronome.setInterval(() => {
  simulation.eventsPer1000Ticks += 100;
}, 1000);
\end{lstlisting}}

Quartermaster provides a mechanism to collect and report information about a simulated execution. Several generally useful metrics are included, such as latencies and event handler counters. It is also possible to extend Quartermaster with additional custom metrics. For example, while simulating the CircleCI incident, we collected latency information and success rates, all of which are included by default, we also added a new metric to collect queue size.

{\small
\begin{lstlisting}
function poll() {
  const queue = service.inQueue as FIFOQueue;
  const queueSize = queue.length();
  stats.record("poll", { queueSize });
}
\end{lstlisting}}

Once a scenario has been defined and the desired metrics recording is in place, a simulation can be executed as follows, where, in this case, $service$ is the entry point for events (i.e., the stage in the model that events are sent to first) and 50,000 is the number of uniformly spaced events that are dispatched.

{\small
\begin{lstlisting}
await simulation.run(service, 50000);
\end{lstlisting}}

\section{Evaluation}
\label{evaluation}

\begin{table*}[t]
\centering
  \caption{Summary of the 8 incidents used in our evaluation.}
  \label{tbl:incidents}
  
  \begin{tabular}{
  r
  l
  >{\raggedright\arraybackslash}p{2.57in}
  >{\raggedright\arraybackslash}p{2.57in}
  r
  }
  \toprule
  
  {\bf \#} &
  {\bf URL} &
  {\bf Architecture} &
  {\bf Failure} & 
  {\bf LOC} \\
  \midrule
  
   1 & \link{bit.ly/2xTn1iK} &
   Load balancer with autoscaled hosts and a database &
   Failed database connections; terminated hosts & 100
  \\ \addlinespace[.2em]
  
   2 & \link{bit.ly/2UNWBrI} & 
   Multiple clients and a shared server & 
   Memory leak and high CPU utilization on server & 70
  \\ \addlinespace[.2em]
  
   3 & \link{bit.ly/2Vku38n} & 
   Redis clusters with primary and secondary instances & 
   Degraded service and cluster nodes failures & 119
  \\ \addlinespace[.2em]
  
   4 & \link{bit.ly/3aY3Hj5} & 
   Third-party library for parsing HTML & 
   Buffer overrun while preparing some HTTP responses & 42
  \\ \addlinespace[.2em]
  
   5 & \link{bit.ly/2yPjFOl} & 
   Job processing application with several databases & 
   A database went into safety shutdown mode & 63
  \\ \addlinespace[.2em]
  
  
   
   19 & \link{bit.ly/3nTXFGW} & 
   Configuration system for app and caching servers & 
   Terminated app and caching servers & 89
  \\ \addlinespace[.2em]
  
   20 & \link{bit.ly/3iYQl9z} & 
   Database clusters with one master and two replicas & 
   Multiple databases were terminated and erased & 94
  \\ \addlinespace[.2em]

   27 & \link{bit.ly/341S9Ku} & 
   Event queue, event processor, and database & 
   Overwhelmed database led to stall in event processing & 76
  \\
  
  \bottomrule
  \end{tabular}
  \vspace{-0.2in}
\end{table*}

As a partial evaluation of Quartermaster, we have used it to model eight systems (or relevant parts of systems) and to simulate publicly reported failures of those systems.\footnote{https://github.com/BYU-SE/idb-quartermaster-models} We selected the eight failures from a database of publicly published incident reports.\footnote{https://github.com/BYU-SE/idb} The goal of this modeling exercise is to demonstrate that (1) the abstractions provided by Quartermaster (primarily stages and events) allow us to model a range of interesting and real systems, and (2) simulated executions of those models appropriately reflect the real system's behavior, in particular its degraded behavior or failures. 

Table~\ref{tbl:incidents} summarizes the eight incidents that we have used in our evaluation. As shown, in the table, we have been able use Quartermaster to model a wide range of architectural components, such as an autoscaling group (incident 1), an in-memory data structure (incident 3), a variety of databases (incidents 1, 5, 19, 20, 27), a load balancer (incident 20), and a plugin for a web server (incident 4). We have also been able to model a wide range of fault-tolerant techniques, such as retries (incident 27), timeouts (incidents 1, 19), caching (incidents 3, 19), and queuing (incidents 1, 3, 19, 20, 27).

These incidents describe failures such as underscaled servers (incidents 1, 27), failed hosts (incidents 3, 5, 19, 20), database failures (incidents 1, 5, 20), and resource leaks (incidents 2, 4). These failures included fail-silent failures (a failed component produces no output or a clearly indicated failure) and Byzantine failures (the failed component continues to run but produces incorrect results)~\cite{Brasileiro_1996}.

While the above modeling exercise has allowed us to validate that Quartermaster can effectively model real systems and simulate real system executions, a more complete evaluation would need to also consider how effectively insights gained through Quartermaster explorations can be applied in a real system and the value of those insights to engineers or researchers. We also hope to evaluate the usability of Quartermaster, and use the results of that evaluation to guide documentation and other improvements. We plan to conduct this more complete evaluation in the future.

\section{Conclusion and Future Work}
\label{conclusion}

Quartermaster is a new modeling and simulation tool that allows engineers and researchers to explore (existing and novel) fault-tolerant techniques and their effects on a system's behavior under different scenarios. Quartermaster distinguishes itself from other system simulation software with a narrow scope to help users understand how and why a system degrades rather than focusing on resource utilization, resource scheduling, or performance. In our partial evaluation, we have demonstrated that Quartermaster can model realistic systems and simulate the behavior of those systems under various scenarios. And we argue that Quartermaster provides these insights with minimal effort for the researcher or engineer.

There are several next steps in this research. We plan to experiment with automatically converting infrastructure definitions, used with infrastructure management software like CloudFormation\footnote{https://aws.amazon.com/cloudformation/} or Terraform,\footnote{https://www.terraform.io/} into an appropriately abstracted Quartermaster model. We will extend Quartermaster with additional prebuilt stages and investigate how we can incorporate quality of service (QoS) metrics for a more targeted analysis of system behavior\cite{Iyengar_2005, Menasce_2002}. Lastly, we hope to use Quartermaster to invent novel techniques that perform well in circumstances where existing techniques do poorly.

\bibliography{IEEEabrv,paper}{}
\bibliographystyle{IEEEtran}

\end{document}